%% file: main.tex
\documentclass[preprint]{article}

\usepackage{microtype}
\usepackage{graphicx}
\usepackage{subcaption}
\usepackage{booktabs} 

\usepackage{hyperref}

\usepackage{icml2026}

\usepackage{amsmath}
\usepackage{amssymb}
\usepackage{mathtools}
\usepackage{amsthm}
\usepackage{tabularx}
\usepackage{makecell}
\usepackage{subcaption}
\usepackage{listings}
\usepackage{minted}
\usepackage{array}
\usepackage[minted,most]{tcolorbox}
\usepackage{multirow}

\usepackage[capitalize,noabbrev]{cleveref}

\theoremstyle{plain}

\theoremstyle{definition}

\theoremstyle{remark}

\usepackage[textsize=tiny]{todonotes}

\lstdefinestyle{prompt}{
  basicstyle=\ttfamily\footnotesize,
  columns=fullflexible,
  breaklines=true,
  frame=single,
  numbers=none,
  numberstyle=\tiny,
  xleftmargin=2em,
  framexleftmargin=1.5em,
  aboveskip=0.6em,
  belowskip=0.6em
}

\icmltitlerunning{Improve Code Reasoning via Concept Learning}

\begin{document}

\newcommand{\mycomment}[1]{}
\newcommand{\tool}{{ConceptCoder}\xspace}
\newcommand{\wei}[1]{{\color{red}Wei:~[#1]}}
\newcommand{\rasel}[1]{{\color{blue}Rasel:~[#1]}}

\twocolumn[
  \icmltitle{ConceptCoder: Improve Code Reasoning via Concept Learning}


  \icmlsetsymbol{equal}{*}

  \begin{icmlauthorlist}
    \icmlauthor{Md Mahbubur Rahman}{yyy}
    \icmlauthor{Hengo Tong}{yyy}
    \icmlauthor{Wei Le}{yyy}
  \end{icmlauthorlist}

  \icmlaffiliation{yyy}{Iowa State University, Ames, Iowa, USA}

  \icmlcorrespondingauthor{Md Mahbubur Rahman}{mdrahman@iastate.edu}

  \icmlkeywords{Machine Learning, ICML}

  \vskip 0.3in
]



\printAffiliationsAndNotice{}  

\input{sections/abstract}
\input{sections/intro}
\input{sections/approach}
\input{sections/evaluation}
\input{sections/related}

\input{sections/conclusions}


\bibliography{bibliography}
\bibliographystyle{icml2026}

\newpage
\appendix
\onecolumn
\input{sections/appendix}




\end{document}

%% file: sections/abstract.tex


\begin{abstract}
Large language models (LLMs) have shown promising results for software engineering applications, but still struggle with code reasoning tasks such as vulnerability detection (VD). We introduce ConceptCoder, a fine-tuning method that simulates human code inspection----models are trained to first recognize {\it code concepts} and then perform reasoning on top of these concepts. In prior work, concepts are extracted by multi-modal models or LLMs to explain vision and natural language models. Our work is the first that formulate concepts for code. We define {\it code concepts} as human-understandable semantic properties of code and trained models to learn such concepts. Our evaluation shows that such approach significantly improves VD accuracy, from 66.32 to 72.15 (F1) on average over 9 open-source LLMs. ConceptCoder achieves the best VD performance compared to the state-of-the-art (SOTA) baselines, including approaches of fine-tuning SOTA open-source LLMs and prompting the proprietary models of GPT-5.2 and Claude-Opus-4.5. Our approach scales in that the concepts defined based on four types of vulnerabilities benefit general vulnerability datasets with 134 CWEs. We also demonstrated that concept-based fine-tuning can generalize beyond VD and improved branch prediction.  We release our code and datasets at \href{https://figshare.com/s/1decab8232c653b44f71}{https://figshare.com/s/1decab8232c653b44f71}.

\mycomment{
\wei{improve: 
To address this gap, we propose ConceptCoder, which decomposes a complex code reasoning task into multiple intermediate steps and formulates them as code concepts. ConceptCoder can learn these concepts and integrate the
semantic representation of these concepts to the reasoning process, thereby, enhancing the models ability to perform complex code reasoning.
understanding of the steps into the code reasoning task.

and contribution/insights 
a novel concept learning framework designed to enhance the code reasoning capabilities of LLM when fine-tuned for code reasoning tasks. Inspired by concept-based approaches in vision-language models, ConceptCoder introduces the idea of code concepts—fine-grained semantic features critical for reasoning tasks and jointly learns code concepts and reasoning tasks via multi-task learning, allowing the model to integrate semantic understanding into downstream code reasoning. We instantiate this approach on two code reasoning tasks: vulnerability detection (VD) and branch prediction (BP). 
}
Large Language models have shown promise in supporting software engineering, but still struggle with complex code reasoning tasks, which require semantic understanding of code concepts without execution.  Through linear probing, we observe that fine-tuning LLMs on code reasoning tasks alone does not lead to a strong understanding of the related concepts. To address this limitation, we propose ConceptCoder, a novel concept learning framework that guides LLMs to better capture and utilize relevant code concepts during reasoning. Our approach begins by defining task-specific concepts and designing a corresponding concept learning objective. We then jointly train the model on both the concept learning task and the code reasoning task using a multi-task learning framework. This encourages the model to incorporate semantic understanding into its reasoning process. We instantiate this approach on two code reasoning tasks: vulnerability detection (VD) and branch prediction (BP). Experiments with ten state-of-the-art LLMs (up to 8B parameters) demonstrate that while pre-trained models poorly capture code concepts, ConceptCoder significantly improves concept understanding (from 56.95 to to 85.64 F1 for VD,  from 27.32 to 32.23 for BP on average over 10 models ) and code reasoning performance (VD F1 from 66.32 to 72.15; BP F1 from 85.27 to 86.50). Furthermore, ConceptCoder enhances model robustness against semantically preserving code perturbations, with average F1 improvements from 19.05 to 25.35 on average across all VD models. ConceptCoder significantly outperformed the SOTA code reasoning tools TRACED and DeepDFA. We published our code and dataset at \href{https://figshare.com/s/1decab8232c653b44f71}{https://figshare.com/s/1decab8232c653b44f71}

}

\end{abstract}

%% file: sections/intro.tex
\section{Introduction}

Large language models (LLMs) have demonstrated increased capabilities for software engineering tasks, e.g., code generation~\cite{code_gen_1, code_gen_2, code_gen_3, code_gen_4}. However, LLMs still face significant challenges of code reasoning tasks such as vulnerability detection (VD). Using standard fine-tuning (SFT), LLMs reported 21.43 F1 on real-world VD datasets~\cite{PrimeVul}. When applied the best prompting techniques from ~\cite{prompting} to the real-world setting of imbalanced vulnerable and non-vulnerable datasets, GPT-5.2 and Claude-Opus-4.5 reported 48.57 and 46.58 F1 respectively. Code reasoning is challenging because the models need to predict program behaviors from code text only. For VD, it required to reason about the safety properties of code. For branch prediction (BP), the models need to evaluate the conditions in the branches. Prior results~\cite{prompting,PrimeVul} indicate that such semantic properties are hard to emerge from code text tokens. 

To improve models' knowledge on code semantics, DeepDFA~\cite{deepdfa} incorporated domain-inspired program analysis algorithm when designing embedding and training. TRACED~\cite{traced} and FuzzPretrain~\cite{fuzzpretrain} pre-trained models with execution traces and input/output, while SemCoder~\cite{semcoder}, CODEI/O~\cite{codeio}, Cycle~\cite{cycle}, and SelfPiCo~\cite{selfpico} fine-tune models using input/output and execution feedback to steer generation or reasoning. Nova~\cite{nova} designed hierarchical attention and contrastive objectives for assembly code, and VulnSC~\cite{vulnc} used LLMs to synthesize inter-procedural semantic data for downstream fine-tuning.

In this work, we introduce {\it code concepts} to help models learn code semantics. In prior work~\cite{corina, srivastava2024vlgcbm, yang2023language, kim2018interpretability}, {\it concepts} are human-understandable semantic features summarized from image pixels or natural language (NL) paragraphs. Concepts are extracted via multimodal-models or LLMs~\cite{clip, open_clip, meta_clip, alpha_clip} and integrated in {\it Concept Bottleneck Models (CBMs)}~\cite{koh2020concept, losch2019interpretability, oikarinen2023label} to improve intepretability of vision and NL models.  To the best of our knowledge, our work is the first that introduced concepts for code and developed concept-based fine-tuning to improve reasoning tasks.

\begin{figure}
    \centering
\includegraphics[width=\linewidth]{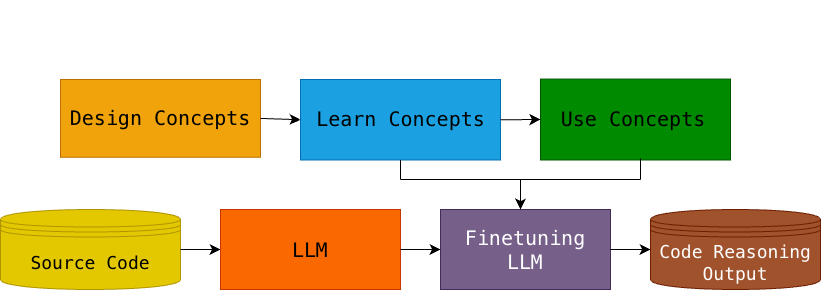}
    \caption{Concept-based LLM fine-tuning: an overview}
    \label{fig:overview}
\end{figure}

We define code concepts as semantic features needed by code reasoning tasks. For VD, we formulate {\it vulnerable concepts} from root cause statements of a vulnerability. For BP, we define concepts as abstract values of statements. Our design ensures that such features are at a proper level, where they can be automatically and accurately computed by domain tools. We developed ConceptCoder (Figure~\ref{fig:overview}), and it first fine-tuned models using code concepts as supervised labels and then conducted a multi-task learning to integrate the final target of code reasoning. This idea is aligned with recent successes in hierarchical reasoning~\cite{multi_step_1, multi_step_2, multi_step_3, multi_step_4, boostrapping_1, boostrapping_2, bootstrapping_3, synthetic}, where we teach models to learn step by step towards complex reasoning goals. Some work focused on COT prompting~\cite{multi_step_1, multi_step_2, multi_step_3, multi_step_4} and others used bootstrapping and learned from small synthetic examples~\cite{boostrapping_1, boostrapping_2, bootstrapping_3, synthetic}. We have also tried COT to prompt models to first learn code concepts and then predict end reasoning tasks; however, we got similar conclusions as~\cite{prompting} and found that COT prompt can't sufficiently address the challenge (Appendix~\ref{cot_prompt}). 

We evaluated ConceptCoder on VD and BP two code reasoning tasks. We used the SOTA real-world VD datasets of PrimeVul \cite{PrimeVul} and \cite{DiverseVul} and constructed a {\it general} dataset covering 80,204 examples and 134 CWE types, as well as a {\it concept} dataset of 28,974 examples, covering four vulnerability types where our concepts are derived from. We fine-tuned 9 open-source LLMs up to 8B parameters (the largest models our computing facilities can support), covering different LLM families such as Llama, StarCoder, Qwen, and  Magicoder. 

Our results show that ConceptCoder significantly improved VD performance over standard SFT, and the average F1 increases from  66.32 to 72.15 in the concept dataset and from 55.11 to 58.52 in the general dataset. ConceptCoder with QN-7B achieves the SOTA performance for VD (74.76 F1 in the concept dataset and 60.6 F1 in the general dataset ), outperforming the best SOTA baselines we know, including DeepDFA~\cite{deepdfa} and TRACED~\cite{traced}, as well as prompting the best LLMs to date, GPT-5.2 and Claude Opus 4.5. Our analysis found that the better the models recognized the concepts, the more VD performance improves; the greater number of concepts learned, and the better VD performance achieves. We also demonstrated that ConceptCoder can generalize over BP and consistently achieved best performance over SFT for all the 9 models: QN-7B reaches 89.37 F1, exceeding TRACED (86.31) and the best standard supervised fine-tuning LLM baseline (88.31).



Our research contributions are as follows:
\begin{enumerate}
\item We define code concepts as statement-level code semantics that can capture intermediate reasoning signals, propose learning them via explicit supervision.
\item We present ConceptCoder, a concept-
supervised multi-task fine-tuning framework that
trains LLMs to  jointly (i) recognize code concepts and (ii) predict end-task labels, enabling
structured semantic reasoning. 
\item We conducted a comprehensive evaluation that demonstrated a significant performance improvement for the VD code reasoning task, and showed our techniques are generally applicable to BP.
\item We release our datasets, code, the tool to support further research in concept-driven code reasoning.
\end{enumerate}

%% file: sections/approach.tex
\section{ConceptCoder}

\subsection{What are code concepts}


A concept is a human-understandable, semantically meaningful property (semantic feature) of an input. It is an intermediate description extracted from low-level features and can be used to explain high-level predictions. For example, given an image of a car, a vision language model can extract from pixels, the concepts of "red" and "four wheels", which is useful to explain why the prediction is a {\it car}~\cite{corina, srivastava2024vlgcbm, yang2023language, kim2018interpretability}. Similarly, in the NL domain, given a sequence of text tokens, LLMs can extract the concepts of "over-priced" and "bad service", which explained why the model predicted a review as {\it negative}~\cite{nlp_concept_1, nlp_concept_2}. 


More generally, a concept is some discrete or continuous value (e.g., binary, categorical, or graded) mapped from a raw input, and it can represent attributes, relations or any abstract/latent factors of interest that align with human reasoning. To be useful, concepts are typically defined as more abstract than raw features (pixels, tokens, embeddings) but less task-specific than the final label.    The notion is modality-agnostic, and  extending it to source code, we define {\it code concepts} as human-understandable program properties extracted from code text. They can be about code structure (loops/recursion), runtime property (control/data flow), behavior (e.g., sorting/searching), resource/API usage (I/O, network, database), safety/security risks (e.g., bounds check, authorization) and design/style (modularity/design patterns). To enable learning concepts for code reasoning, we defined concepts as certain semantics of code that can serve as an intermediate step of reasoning and be computable by domain tools so that we can obtain supervised labels.


Specifically, for VD, we review domain knowledge of code inspection and human-constructed program analysis tools, and identified seven {\it vulnerable concepts} from four frequently occurred vulnerability types, including {\it memory leak}, {\it use-after-free}, {\it buffer overflow} and {\it null-pointer dereference}. For example, to detect memory leaks, we need to know whether the allocated memory is released along all program paths; therefore, recognizing concepts of "memory allocation" and "memory release" is an important step of detecting this vulnerability. We designed a mapping from the type of a statement to the 7 code concepts we modeled (Table~\ref{tab:concept_map_vd}), and 
 developed a static analysis tool to compute such concepts as labels for concept learning. This approach promotes the models to learn vulnerability-indicative patterns over these concept sequences instead of directly from code tokens.

\begin{table}[ht]
\centering
\caption{Code Concepts for VD: Mapping Code Token Sequence to Vulnerability Concepts}
\resizebox{0.45\textwidth}{!}{ 
\begin{tabular}{l|r}
\Xhline{1.5\arrayrulewidth}
\textbf{Token Sequence Example} & \textbf{\makecell{Concept for VD \\ }} \\
\Xhline{1.5\arrayrulewidth}
\texttt{p = NULL} & Null assignment \\ \hline
\texttt{if (p == NULL)} & Null check \\ \hline
\texttt{p->dims = 2} & Pointer dereference \\ \hline
\makecell[l]{\texttt{p=(char *)malloc(total+1)} \\ \texttt{char p[4]}} & Memory allocation \\ \hline
\texttt{p[len] += 1} &\makecell[r]{Buffer access \\Pointer dereference} \\ \hline
\texttt{if (i <= strlen(p))} & Bounds check \\ \hline 
\texttt{free(p)} & Memory free \\
\hline
\Xhline{1.5\arrayrulewidth}
\end{tabular}
}
\label{tab:concept_map_vd}
\end{table}

\subsection{Learning Code Concepts}

The goal of concept learning is to train a model to recognize the code concepts from code text. Given a function, we split it into a sequence of statements. For each statement $s$, we assign a multi-hot concept label
$\mathbf{y}\in\{0,1\}^{N}$, where $N$ is the number of concepts supported by our framework, and $y_i=1$ indicates the presence of concept $i$ in $s$ (otherwise $y_i=0$).

For training, we feed the model’s statement representation into a classification head that outputs logits in $\mathbb{R}^{N}$; a sigmoid converts them to probabilities $\mathbf{p}\in(0,1)^N$. Since multiple concepts may co-occur in the same statement, we formulate concept prediction as multi-label classification and optimize Binary Cross Entropy (BCE):
\begin{equation}
\mathcal{L}_{\text{concept}} = -\frac{1}{N} \sum_{i=1}^{N} \left[ y_i \cdot \log(p_i) + (1 - y_i) \cdot \log(1 - p_i) \right],
\end{equation}
where $y_i \in \{0,1\}$ is the ground-truth label for concept $i$, $p_i \in (0,1)$ is the predicted probability after sigmoid, and $N$ is the number of concepts.



\subsection{Using Concepts in Code Reasoning via Multitask learning}

\begin{figure}
    \centering
\includegraphics[width=\linewidth]{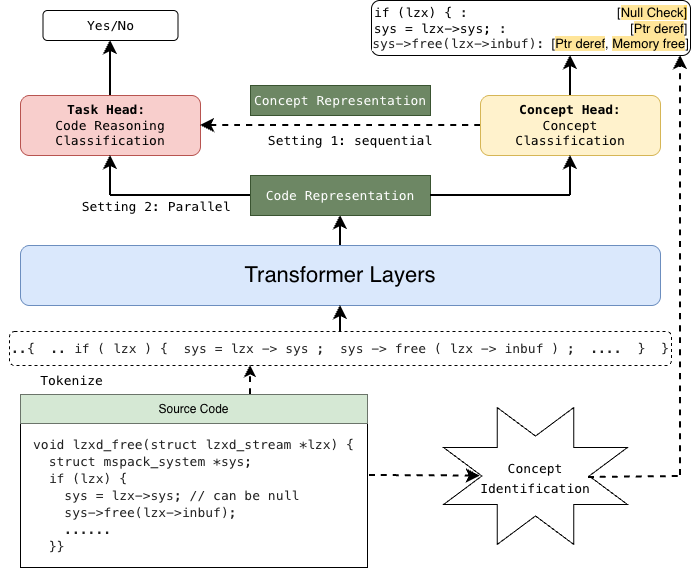}
    \caption{Supervised Multi-Task Learning: Joint Training for Code Concepts and Code Reasoning}
    \label{fig:methodology}
\end{figure}

We incorporate concept learning into code reasoning via multi-task learning by training a shared encoder with two heads: a concept head and a task head (Figure~\ref{fig:methodology}). The concept head predicts a concept probability vector $\mathbf{p}\in(0,1)^N$ for every statement, while the task head predicts the downstream task label.

We study two integration strategies. In the \textit{sequential} setting, the task head consumes the concept predictions (e.g., aggregated across statements), so the final decision is explicitly guided by the learned concepts. In the \textit{parallel} setting, both heads extract the input from the shared transformer layers and are optimized jointly; concept supervision regularizes the encoder, while the task head uses the encoder features directly. The overall training objective is a weighted sum of the concept loss and VD loss:
\begin{equation}
\mathcal{L}_{\text{total}} = \lambda_1 \mathcal{L}_{\text{concept}} + \lambda_2 \mathcal{L}_{\text{VD}},
\end{equation}
where $\lambda_1$ and $\lambda_2$ control the relative contribution of each task.

In our vulnerability detection experiments, we reported results using the sequential setting as it performed better.

\subsection{ConceptCoder for Other Code Reasoning Task}
To demonstrate that our concept-learning framework can generalize beyond VD, we additionally instantiate it for branch prediction (BP), where the goal is to predict whether a branch will be taken for a given input. BP requires reasoning about runtime values of variables and expressions that control branching. Therefore, we define code concepts for BP by mapping code token sequences to \emph{abstract values} they compute. These concepts form an interpretable intermediate representation of value-flow signals relevant to branch behavior. This idea is inspired by prior work~\cite{traced,trace-pradel}, where researchers demonstrated that models cannot learn from exact numeric values, but can learn from quantized values abstracted from execution traces.

\label{app:bp_concepts}
\begin{table}[ht]
\centering
\caption{Code Concepts for BP: Mapping Code Token Sequences to Abstract Values }
\resizebox{0.48\textwidth}{!}{ 
\begin{tabular}{l|r}

\Xhline{2\arrayrulewidth}
\Xhline{1.5\arrayrulewidth}
\textbf{Token Sequence Example} & \textbf{\makecell{Concept \\  for BP}} \\
\Xhline{1.5\arrayrulewidth}
\makecell[l]{\texttt{x > y} \\ \texttt{x == y} \\ \texttt{other conditions}} & True, False \\ 
\hline
\texttt{x = 'M'} & Alphabetic \\ \hline
\texttt{x = '='} & Non-alphabetic  \\ \hline
\texttt{x[0] = 'a';}  & Initialized \\ \hline
\texttt{char *x = strchr(a + 1, '0');} & NULL, Not NULL \\ \hline 
\texttt{x = a * b + c} & \makecell[r]{Zero \\ Positive regular \\ Positive large \\ Negative regular \\ Negative large}  \\ \hline 
\Xhline{1.5\arrayrulewidth}
\end{tabular}
}
\label{tab:concept_map_bp}
\end{table}

In Table~\ref{tab:concept_map_bp}, we present 12 concepts for BP together with code examples. The mappings are automatically computed by our static analysis tool, following our algorithm of mapping concrete values of a statement to the corresponding abstract values based on the statement type. For example, the first row of Table~\ref{tab:concept_map_bp} indicates that boolean expressions are mapped to an abstract value of {\tt True} or {\tt False}. We first ran our training copra, collected their execution traces and obtained concrete values, from which, we then computed abstract values as supervised labels for concept learning.

In multi-task learning, we found that parallel setting performs better for BP. Specifically, the concept head predicts the abstract values of the statements, while the classification head receives a representation from tokens of a specific branch and predicts whether that branch is taken for a particular input. See Appendix  \ref{app:app_bp_dataset} for further details.

\subsection{Concept Probing}
\label{concept_probing}
To determine if code representations have encoded designed code concepts, we followed the literature~\cite{linear_probing} and trained a linear classifier that takes in code representations from fine-tuned LLMs and outputs a vector  of code concepts, {indicating the presence of each concept}. The training dataset of the concept probing model is distinct from the ones used for LLM fine-tuning to avoid data leakage (Appendix~\ref{app:concept_probing_dataset}).  Independent of different settings of multi-task learning, concept probing takes code representations {from transformer layers} after concept learning and VD learning are both done.

%% file: sections/evaluation.tex
\section{Evaluation}
\label{sec:results}
\paragraph{Implementation.} We implemented ConceptCoder by fine-tuning the SOTA LLMs, following the practices from prior work \cite{PrimeVul, DiverseVul, codexglue}.  We used a learning rate of $2 \times 10^{-5}$ and the Adam optimizer. Training is conducted for 10 epochs with a batch size of 64. We fine-tune the models under the bf16 data types. We developed static analysis tools to compute the concepts. 

\paragraph{Datasets.} We used the SOTA vulnerability datasets of PrimeVul and DriverseVul~\cite{DiverseVul, PrimeVul} and constructed a {\it general} dataset, consisting of 80,204 examples and 134 CWE types. We also built a {\it concept} dataset, a total 28,974 examples, consisting of four vulnerability types including null-pointer dereference, buffer overflow, memory leak, and use-after-free. We keep the datasets imbalanced to mimic the real-world settings. For BP, we used the CodeNet dataset~\cite{codenet} and processed it using Traced \cite{traced} Tools. We obtained a total dataset of 52060 examples (Appendix \ref{app:app_bp_dataset}). We split each dataset into training, testing, and validation using an 80:10:10 ratio. We used F1 as a metric, following the literature~\cite{CausalVul, traced, PrimeVul}.




\paragraph{Models.}

\begin{table}[]
\caption{Models and their IDs in the Tables and Figures.}
\resizebox{0.48\textwidth}{!}{ 
\begin{tabular}{lll}
\hline
 Model & \makecell[l]{Full Model Name \\ in HuggingFace} & \makecell[l]{Model IDs } \\ \hline\hline
 Qwen2.5-3B & Qwen/Qwen2.5-Coder-3B  & QN-3B \\
 Qwen2.5-7B & Qwen/Qwen2.5-Coder-7B & QN-7B \\
 StarCoder2-3B & bigcode/starcoder2-3b  & SC-3B \\
 StarCoder2-7B & bigcode/starcoder2-7b & SC-7B \\
 Llama3.2-3B & meta-llama/Llama-3.2-3 & LM-3B \\
 Llama3.1-8B & meta-llama/Llama-3.1-8B & LM-8B \\
 CodeLlama-7B & meta-llama/CodeLlama-7b-hf & CLM-7B  \\
Magicoder-7B & ise-uiuc/Magicoder-S-DS-6.7B  & MC-7B \\
 CodeGemma-7B& google/codegemma-7b & CG-7B \\ \hline
\end{tabular}
}
\label{tab:models}
\end{table}
We applied our approach to 9 pre-trained LLMs, shown in \autoref{tab:models}.
These are the SOTA models trained on code data and frequently used for software engineering tasks
\cite{llmstudy,llmsurvey, semcoder,cycle,codeio,llm1,llm2,llm3,llm4,llm5}. 8B is the largest models we can fine-tune in our computing facilities. We fine-tuned classification heads from the base models, not the chat or instruction versions. 



\subsection{ConceptCoder Outperformed SFT and Scales to Diverse Types of Vulnerabilities}

\paragraph{Setup.}  We first evaluated our effectiveness on the concept dataset which contains the same four vulnerability types where our concepts are derived. We then investigated whether ConceptCoder can scale to the general dataset, and benefit a diverse set of vulnerability types, even though concept training is based only on the same seven concepts. We compared ConceptCoder with standard supervised fine-tuning (SFT)~\cite{PrimeVul}. The two approaches used the same training and testing data from concept and general datasets respectively in two settings. 


\paragraph{Results.}

\begin{figure*}[t]

    \centering
    \resizebox{0.94\textwidth}{!}{ 
    \begin{subfigure}[t]{0.48\textwidth}
        \centering
        \includegraphics[width=\linewidth]
        {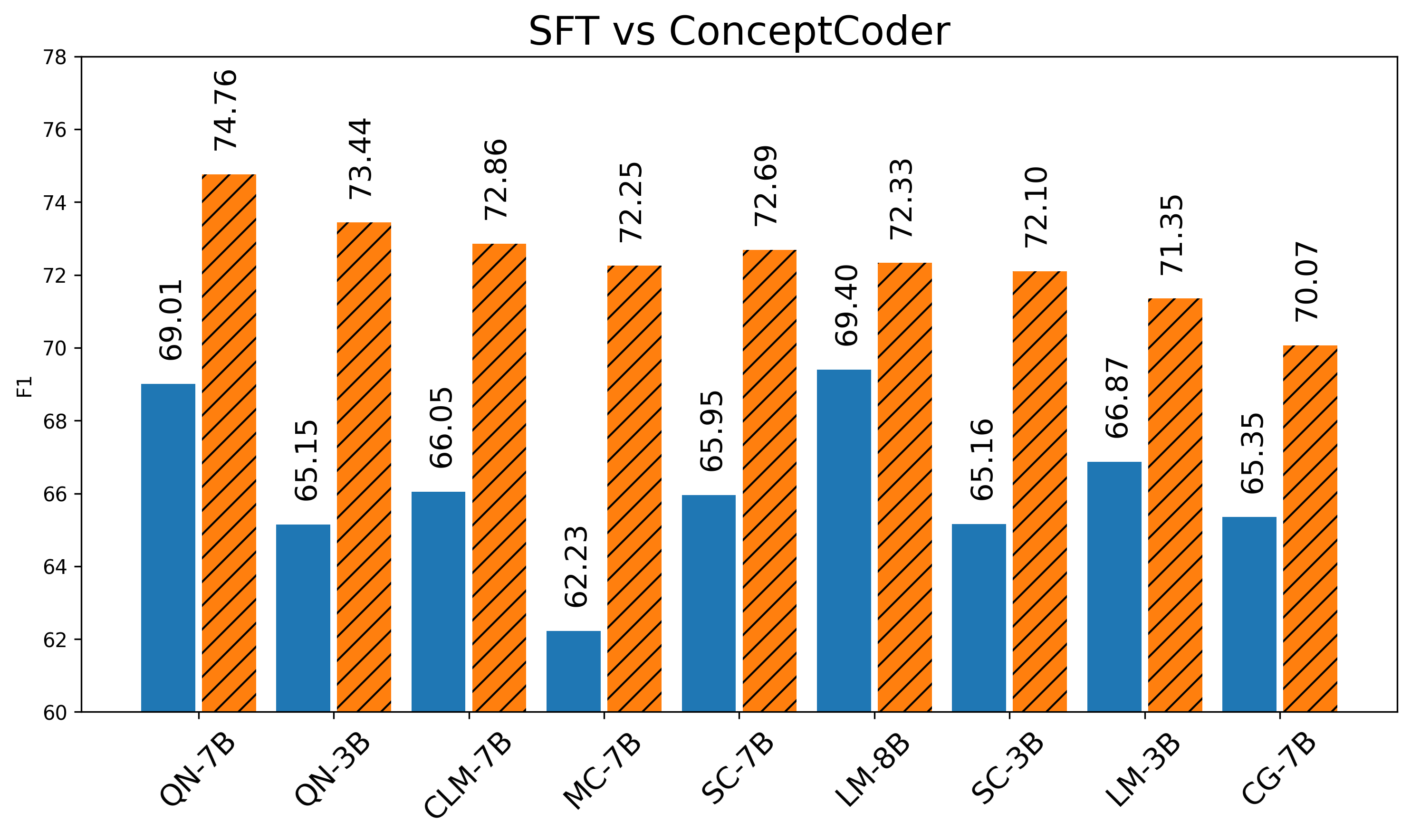}
        \caption{ConceptCoder outperformed SFT on {\it the concept} dataset.}
            \label{fig:rq1-result1}
    \end{subfigure}
    \hfill
    \begin{subfigure}[t]{0.48\textwidth}
        \centering
        \includegraphics[width=\linewidth]{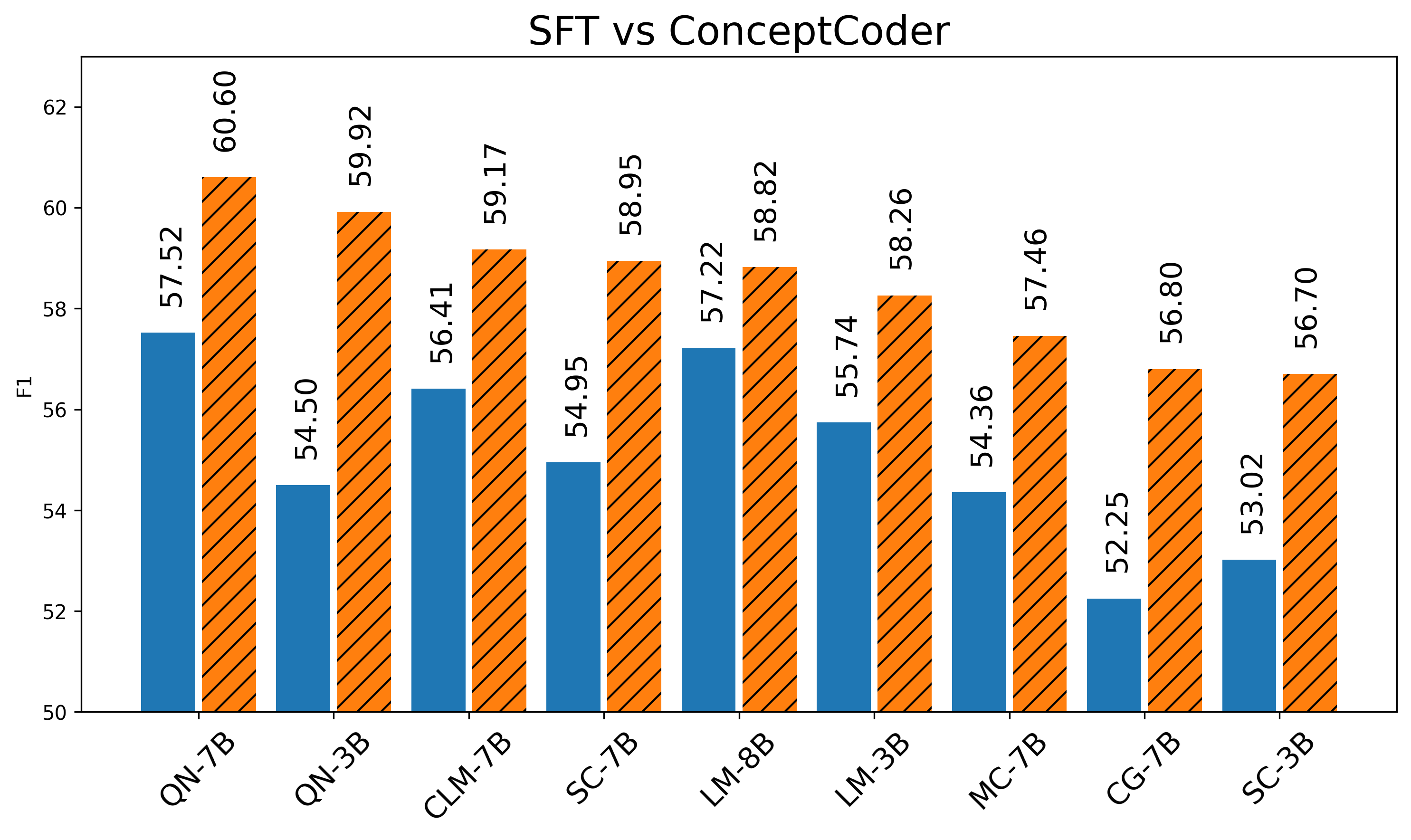}
        \caption{ConceptCoder scales to the {\it general} dataset}
            \label{fig:rq1-result2}
    \end{subfigure}
    }

    \caption{ConceptCoder significantly outperformed SFT and it benefits the vulnerability types beyond concept data}
    \label{fig:rq1-result}

\end{figure*}

ConceptCoder significantly outperformed SFT (Fig.~\ref{fig:rq1-result1}) on the concept dataset, raising F1 for all the 9 models. The average F1 is improved from 66.32 to 72.15. It achieved the best F1 of 74.76 with Qwen2.5-7B~(QN-7B) and the maximum F1 increase for Magicoder-7B~(MC-7B), from 62.23 to 72.25. 

Interestingly, ConceptCoder scaled to the general dataset (Fig.~\ref{fig:rq1-result2}), raising average F1 across nine models from 55.11 to 58.52. It achieved the best F1 of 60.6 with Qwen2.5-7B (QN-7B), increases Qwen2.5-3B (QN-3B) F1 most from 54.50 to 59.92.  We observed that some concepts defined based on four types of vulnerabilities are also useful for detecting other types of vulnerabilities. For example, knowing where is the bound check helped CWE-20 (Improper Input Validation). 

Different from SFT, ConceptCoder trained the models to first recognize relevant semantics of statements (which is an important step of vulnerability reasoning) ----Fig.~\ref{fig:vd_concept_recog} (Appendix \ref{concept_probing_performance}) showed that the concept learning has significantly improved concept recognition for both datasets. Our results imply that teaching models think step by step helped for complex code reasoning, and we don't need to exhaustively find all the vulnerable concepts to make ConceptCoder useful.











\subsection{ConceptCoder Achieved the Best among SOTA}
\begin{table}[]
    \centering
    \renewcommand{\arraystretch}{1.15}
    \caption{ConceptCoder significantly outperformed the SOTA tools}
    \resizebox{0.48\textwidth}{!}{ 
    \begin{tabular}{l|c|c}
    \hline
        \textbf{Model} & \textbf{Concept Dataset} & \textbf{General dataset}  \\ \hline \hline

       DeepDFA + UniXcoder   & 66.13 & 53.05 \\ 
        DeepDFA + QN-7B   & 69.74 & 58.03\\ 
        TRACED & 67.21 &  52.45 \\ 
        Best SFT LLM  & 69.40 (LM-8B) & 57.52(QN-7B) \\ 
        \hline
        Qwen2.5-Coder-32B-Instruct & 25.11 & 26.65 \\ 
        Llama-3.3-80B-Instruct & 35.35 & 35.85\\ 
        GPT-5.2 & 46.58 & 37.33\\
        Claude-Opus-4.5 & 48.57 & 42.10\\
        \hline\hline
        Best ConceptCoder LLM  & \textbf{74.76}(QN-7B) & \textbf{60.6} (QN-7B)\\ \hline

    \end{tabular}
}
    \label{tab:baselines_sota}
\end{table}



\paragraph{Setup.} We collected the best VD methods we knew and compared with ConceptCoder. The baselines include TRACED~\cite{traced},  DeepDFA~\cite{deepdfa} (two setttings ---- combined with UniXcoder from the original paper, and with Qwen2.5-7B, the best-performing LLM in our list), as well as the best prompting methods for VD (Zero Shot with COT \cite{prompting}) using two larger open-source LLMs, Qwen2.5-Coder-32B-Instruct and Llama3.3-80B-Instruct, and the best proprietary models GPT-5.2 and Claude Opus 4.5. 



\paragraph{Result.} 

Table \ref{tab:baselines_sota} shows ConceptCoder with Qwen2.5-7B (QN-7B) achieves the best performance on both datasets, outperforming all baselines by a clear margin. Among the baselines, DeepDFA + Qwen2.5-7B attains the best F1, reaching 69.74 on the concept dataset and 58.03 on the general dataset. 

Prompting-based methods performed poorly, even with larger open-source models and stronger proprietary models such as GPT-5.2 and Claude Opus 4.5—indicating that VD is a very challenging task and prompting with the best SOTA model is still insufficient. Till today, fine-tuning on vulnerability detection data is necessary for strong performance and that ConceptCoder provides an effective way to do so.


\subsection{Analysis and Additional Findings}


\subsubsection{When concepts learning improved more, VD improves more}

\begin{figure*}[htbp]
    \centering
    \resizebox{0.94\textwidth}{!}{ 
    \begin{subfigure}[t]{0.48\textwidth}
        \centering
        \includegraphics[width=\linewidth]{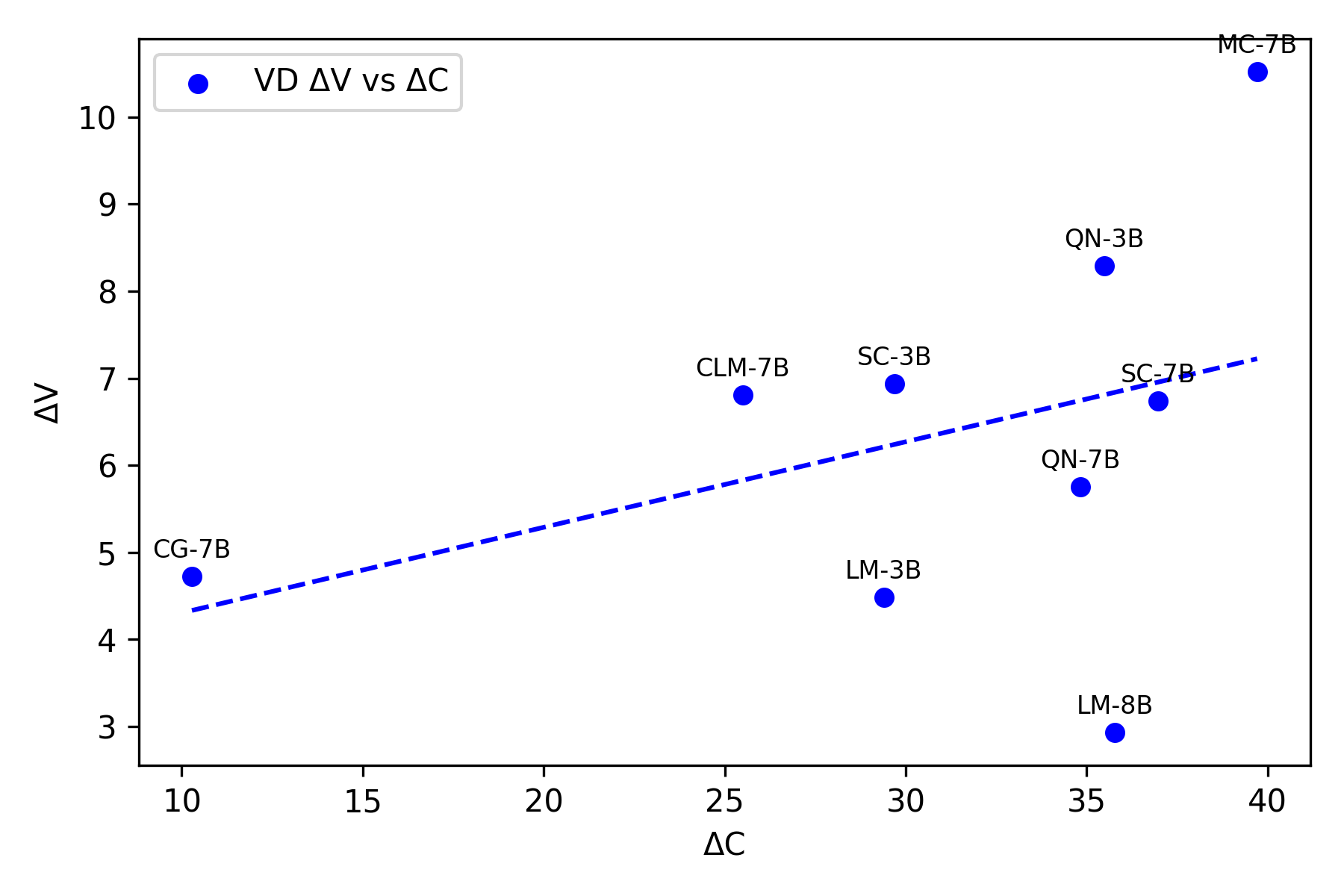}
    \end{subfigure}
    \hfill
    \begin{subfigure}[t]{0.48\textwidth}
        \centering
        \includegraphics[width=\linewidth]{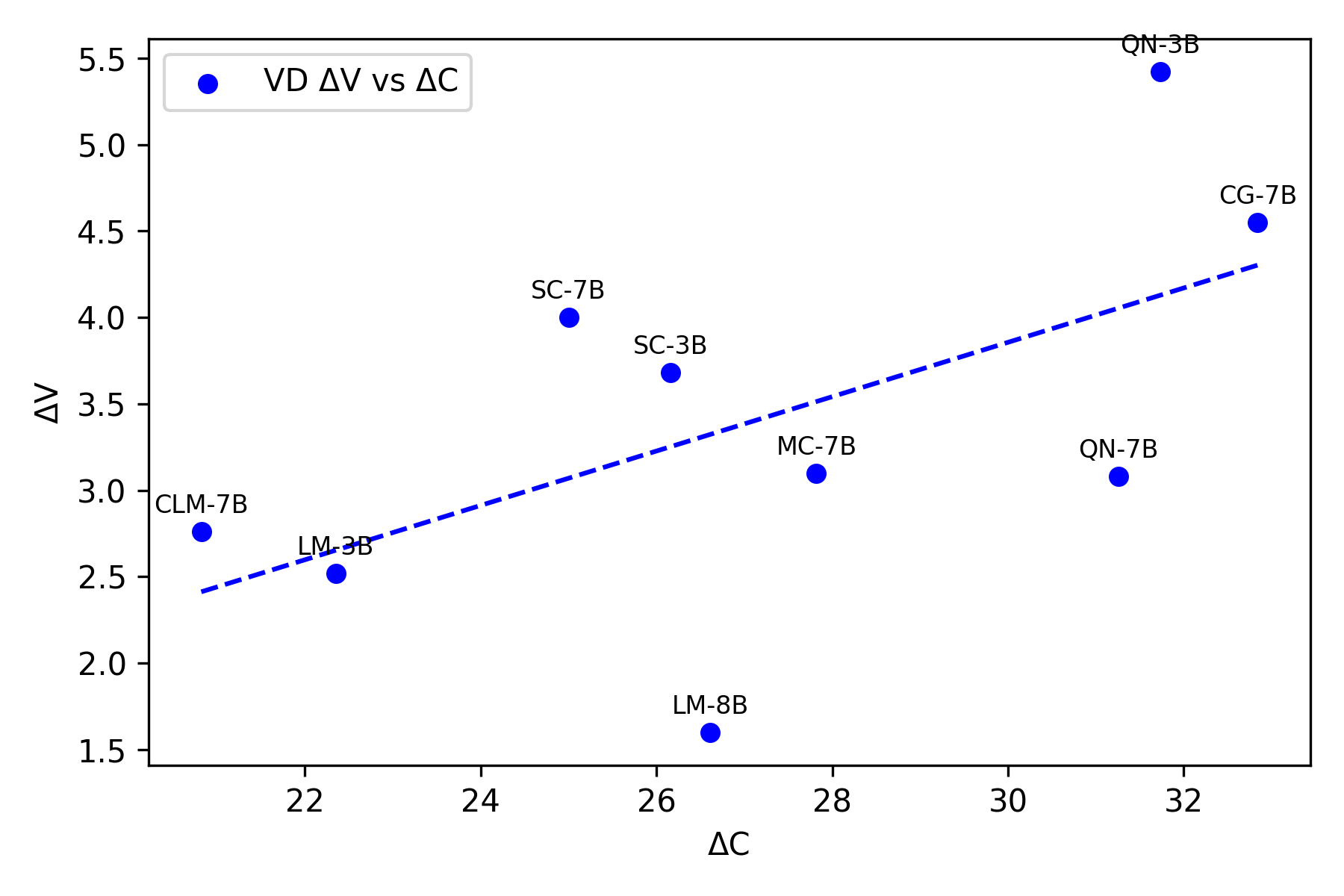}
    \end{subfigure}
    }
    \caption{When the model improved concept understanding more, their VD performance improved more, except LM-8B. Left panel: concept dataset, right panel: general dataset.}
    \label{fig:rq3.1-result}
\end{figure*}

\paragraph{Setup.} For each of the nine models, we measured how well a model learned concepts after ConceptCoder and SFT using our concept probe (Section~\ref{concept_probing}), namely $F1_{CC}$ and $F1_{SFT}$. We computed $\Delta C = F1_{CC} - F1_{SFT}$. Similarly, we obtain $\Delta V$ computed as F1 score difference for VD using two approaches. We plotted the relations of $\Delta V$ and $\Delta C$ for nine models. See Fig.~\ref{fig:rq3.1-result}.



\paragraph{Result.} We observe a general trend for both concept and general datasets----when the model improved concept understanding more, their VD performance improved more. For example, in Fig.~\ref{fig:rq3.1-result} (left), CG-7B shows the smaller improvement in concept learning ($\Delta C=10.28$) and correspondingly smaller VD improvement ($\Delta V =4.72$), whereas MC-7B has the largest $\Delta C$ ($39.71$) and also the strongest VD gain ($\Delta V =10.52$). On the general dataset in Fig.~\ref{fig:rq3.1-result} (right), CLM-7B and LM-3B have relatively low $\Delta C$ ($20.82$ and $22.35$) and only modest VD improvements ($\Delta V = 2.76$ and $2.52$), while models with larger $\Delta C$, such as QN-3B ($31.73$) and CG-7B ($32.84$), achieve noticeably higher VD gains ($\Delta V =5.42$ and $4.55$, respectively).






\subsubsection{When we trained with more concepts, VD improved more} 


\paragraph{Setup.} We investigated whether there is a correlation between the number of concepts learned and the improvement of VD performance. To setup, we trained ConceptCoder 7 times for each model, starting with 1 concept and add one more concept each time (randomly selected from the 7 concept pool) until all the concepts are included. All the models used the same set of 1-7 concepts. This experiment is computationally expensive, and therefore we selected the three best-performing models from Fig.~\ref{fig:rq1-result2} for the study.


\begin{figure}[]
  \centering
\includegraphics[width=0.47\textwidth]{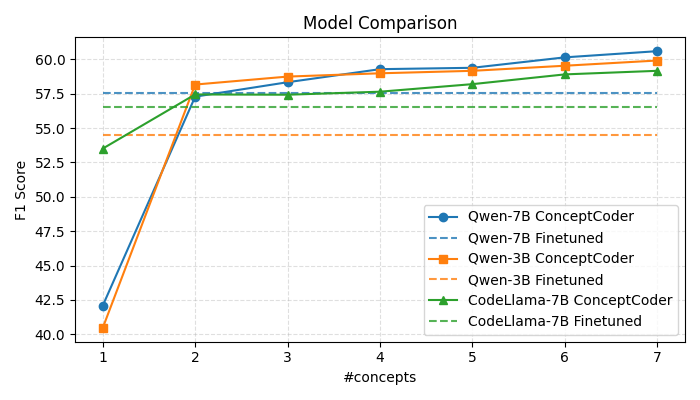}
  \caption{F1 score vs. number of concepts (n) for ConceptCoder across the best three models. Dashed lines indicate the corresponding SFT baselines.}
  \label{n_concept}
\end{figure}

\paragraph{Result.} Fig.~\ref{n_concept} shows that when the number of concepts increased, the models performed better. For all the three models, the ConceptCoder's performance increases significantly when moving from 1 to 2 concepts. After introducing 2 concepts, Qwen-3B and CodeLlama-7B outperformed SFT, and Qwen-7B performed similar to SFT. We also see that some data points improved more, as some concepts may be more informative than the others, or we need two (or more) concepts learned together to be useful for VD.


\subsubsection{ConceptCoder generalized better and improved robustness}

\paragraph{Setup.} We explored other properties of models such as the abilities of generalizing to unseen data and robustness under perturbation. For generalizability, we tested ConceptCoder trained with the concept dataset on the test split of the general dataset (unseen data). For robustness, we followed the literature~\cite{CausalVul} and performed semantic-persevering code transformations for test data, using the NatGen~\cite{natgen} tool. Specifically, we replaced variable names and inserted API calls as dead code. Those variable names and API calls only appeared in the opposite class of the training dataset.

\begin{table*}[t]
\centering
\small
\setlength{\tabcolsep}{8pt}
\renewcommand{\arraystretch}{1.15}
\caption{Generalization (top two rows)  and robustness (bottom two rows) performance of ConceptCoder compared to finetuned without concept learning.}
\begin{tabular}{lccccccccc}
\hline
\textbf{Approach} & \textbf{QN-3B} & \textbf{QN-7B} & \textbf{CLM-7B} & \textbf{SC-3B} & \textbf{SC-7B} & \textbf{LM-3B} & \textbf{LM-8B} & \textbf{MC-7B} & \textbf{CG-7B} \\
\hline

SFT-G  & 48.53 & 47.47 & 47.49 & 46.38 & 48.93 & 48.75 & 48.42 & 45.38 & 49.01 \\
ConceptCoder-G & \textbf{49.45} & \textbf{50.42} & \textbf{49.49} & \textbf{46.55} & \textbf{49.06} & \textbf{50.05} & \textbf{49.01} & \textbf{48.44} & \textbf{49.46} \\

\hline\hline

SFT-R & 11.91 & 17.29 & 30.16 & 6.23 & 15.25 & 19.98 & 28.4 & 24.66 & 11.19 \\
ConceptCoder-R & \textbf{27.09} & \textbf{25.36} & \textbf{32.63} & \textbf{13.00} & \textbf{18.02} & \textbf{31.72} & \textbf{32.27} & \textbf{38.52} & \textbf{18.06} \\ \hline

\end{tabular}

\label{tab:gen_rob}
\end{table*}

\paragraph{Result.}  Table \ref{tab:gen_rob} reported a consistent improvement of ConceptCoder over SFT for every model. Interestingly, Magicoder-7B ({ MC-7B}) demonstrates the most significant performance improvement in the concept dataset (Fig.~\ref{fig:rq1-result1}), and it also achieves the highest overall robustness and generalization improvement among all 9 models.

For robustness, we observed that smaller models ({ QN-3B}, { SC-3B}, {LM-3B}) benefited from concept learning and improved robustness more. With concept learning, LLaMA-based models ({ LM-3B}, { LM-8B}, { CLM-7B}) achieved high robustness performance----their F1 scores are all above 30, while StarCoder family  models ({ SC-3B}, { SC-7B}) face challenges to achieve VD robustness.




\subsubsection{Non-vulnerable Concepts are useful.}

\paragraph{Setup.} We further hypothesized that the patterns of {\it non-vulnerable} concepts may be useful for distinguishing vulnerable and non-vulnerable classes. We excluded the statements that are mapped to vulnerable concepts, and classify the rest of statements into a diverse set of {\it non-vulnerable} concepts, including vulnerability-irrelevant API Calls, arithmetic operations with no pointers, constant assignments, {\tt if} statements, {\tt jump} statements, loop heads, and {switch} cases. In {\it ConceptCoder-NVC}, we trained models to classify a statements into categories that belong to only non-vulnerable concepts. We then compared it to SFT and our previous setting {\it ConceptCoder-VC} in Table~\ref{tab:nvc_vc}.


\begin{table*}[t]
\centering
\small
\setlength{\tabcolsep}{6pt}
\renewcommand{\arraystretch}{1.15}
\caption{Performance of ConceptCoder using vulnerable and non-vulnerable concepts. The top four rows report results on the concept dataset, and the bottom four rows report results on the general dataset. NVC denotes non-vulnerable concepts, VC denotes vulnerable concepts.}
\begin{tabular}{llccccccccc}
\hline
\textbf{Dataset} &\textbf{Approach} & \textbf{QN-3B} & \textbf{QN-7B} & \textbf{CLM-7B} & \textbf{SC-3B} & \textbf{SC-7B} & \textbf{LM-3B} & \textbf{LM-8B} & \textbf{MC-7B} & \textbf{CG-7B} \\\hline
\multirow{3}{*}{\begin{tabular}[c]{@{}c@{}}Concept\\Dataset\end{tabular}} & SFT          & 65.15 & 69.01 & 66.05 & 65.15 & 65.95 & 66.87 & 69.40 & 62.23 & 65.35 \\
&ConceptCoder - NVC & 71.51 & 72.37 & 70.07 & 70.29 & 71.92 & 70.66 & 71.46 & 69.21 & 69.81 \\
&ConceptCoder - VC  & \textbf{73.44} & \textbf{74.76} & \textbf{72.86} & \textbf{72.10} & \textbf{72.69} & \textbf{71.35} & \textbf{72.33} & \textbf{72.75} & \textbf{70.07} \\
\hline
\hline
\multirow{3}{*}{\begin{tabular}[c]{@{}c@{}}General\\Dataset\end{tabular}}
 &SFT          & 54.50 & 57.52 & 56.41 & 53.02 & 54.95 & 55.74 & 57.22 & 54.36 & 52.25 \\
&ConceptCoder - NVC & 58.99 & 59.36 & 58.99 & 56.67 & 58.08 & 56.86 & 57.43 & 56.76 & 56.37 \\
&ConceptCoder - VC  & \textbf{59.92} & \textbf{60.60} & \textbf{59.17} & \textbf{56.70} & \textbf{58.95} & \textbf{58.26} & \textbf{58.82} & \textbf{57.46} & \textbf{56.80} \\

\hline
\end{tabular}
\label{tab:nvc_vc}
\end{table*}


\paragraph{Result.} We found that {\it ConceptCoder-NVC} also outperformed SFT but {\it ConceptCoder-VC} performed better, indicating that 
non-vulnerable concepts are useful but not as useful as vulnerable concepts.

A plausible explanation is that once models extracted the concepts, they can better associate patterns of concepts with labels than purely used tokens in code text. Vulnerable code and non-vulnerable code indeed have some differences in such patterns to some extent. Vulnerable concepts provides a stronger signal.

\subsection{Generalizing Beyond Vulnerability Detection}
\paragraph{Setup.}  We evaluated our approach for branch prediction. We constructed the dataset based on 12 concepts (Table~\ref{tab:concept_map_bp}) and the CodeNet dataset \cite{codenet}. We compared ConceptCoder with standard supervised fine-tuning and the SOTA baseline, TRACED \cite{traced}.


\paragraph{Result.}
\begin{figure*}[t]
   \centering
    \resizebox{0.94\textwidth}{!}{ 
    \begin{subfigure}[t]{0.48\textwidth}
        \centering
        \includegraphics[width=\linewidth]{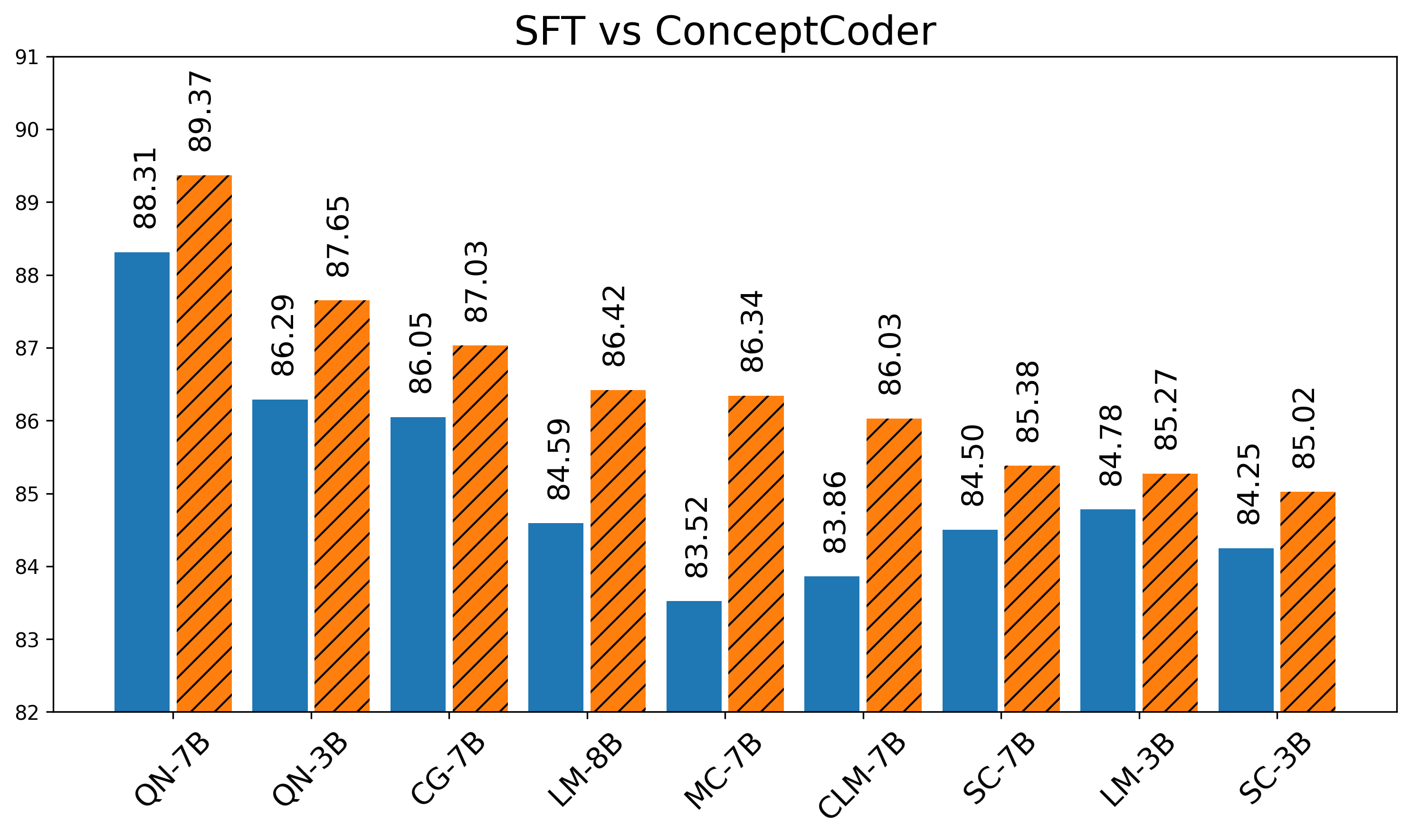}
        \caption{Branch prediction performance}
        \label{fig:bp-result_11}
    \end{subfigure}
    \hfill
    \begin{subfigure}[t]{0.48\textwidth}
        \centering
        \includegraphics[width=\linewidth]
        {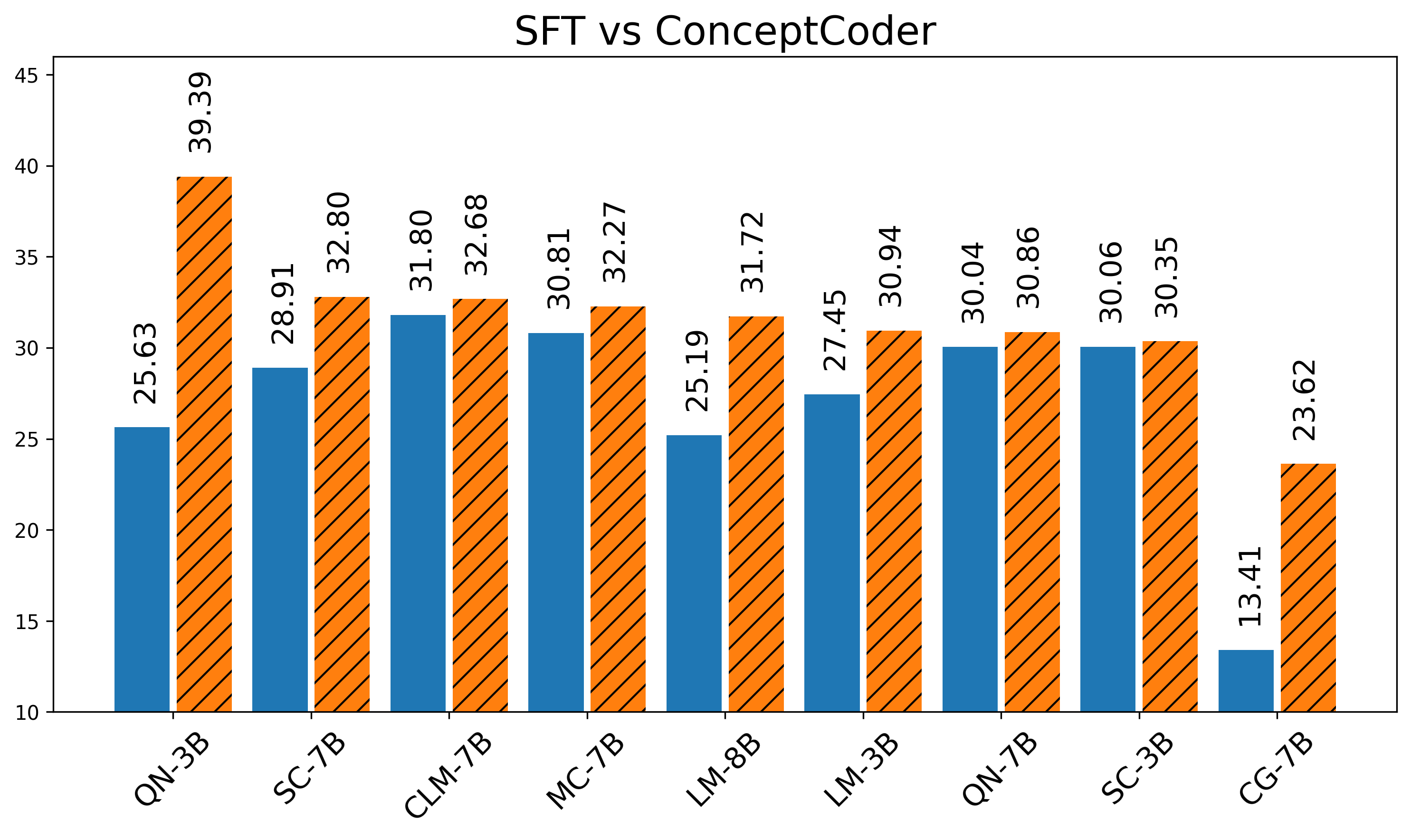}
        \caption{Concept probing performance of branch prediction.}
            \label{fig:bp-result_12}
    \end{subfigure}
    }

    \caption{ConceptCoder significantly outperformed SFT on concept recognition and branch prediction.}
    \label{fig:bp-result_1}
    
\end{figure*}

Figure \ref{fig:bp-result_1} shows that branch prediction also benefits from ConceptCoder for all the models. Across models, average BP F1 increases from 85.27 to 86.50, while concept-recognition F1 improves from 27.75 to 32.25. Models with the largest concept gains (e.g., QN-3B, LM-8B, CG-7B) also tend to see larger BP improvements. Concept recognition F1 remains lower overall because BP concept recognition requires predicting quantized runtime values, which is more difficult than VD concept recognition.

Table~\ref{tab:bp_baselines_sota} reports that ConceptCoder outperformed the baselines: QN-7B reaches 89.37 F1, exceeding TRACED (86.31) and the best SFT LLM QN-7B (88.31). These results show ConceptCoder consistently improves code reasoning and generalizes across both VD and BP. For both tasks, our approach benefits all the models we experimented.

\begin{table}[]
    \centering
        \caption{Comparing with SOTA branch prediction tools}
           \resizebox{0.43\textwidth}{!}{ 
    \begin{tabular}{c|c}
    \hline
        \textbf{Model} & \textbf{F1} \\ \hline \hline
        TRACED - Branch & 86.31 \\ \hline 
        Best fine-tuned LLM - Branch (QN-7B) &  88.31 \\ \hline
        Best ConceptCoder LLM - Branch (QN-7B) & \textbf{89.37} \\ \hline
    \end{tabular}
}
    \label{tab:bp_baselines_sota}
\end{table}



%% file: sections/related.tex
\section{Related Work}
\paragraph{Concepts in vision and NL models.}  
Contrastive Language-Image Pretraining (CLIP) models, such as  OpenAI CLIP \cite{clip}, OpenCLIP \cite{open_clip}, Meta-CLIP \cite{meta_clip}, and ALpha-CLIP \cite{alpha_clip}, learned rich visual concepts by aligning images with natural language descriptions from image–text pairs, enabling zero-shot transfer to diverse vision tasks. 
Building on these capabilities, many works use CLIP-derived concepts to improve or interpret  a wide range of downstream vision tasks \cite{clapp_1, clapp_2,clapp_3, clapp_4, yang2023concept}. 
A parallel line of work focuses on Concept Bottleneck Models (CBMs)  originally developed to improve intepretability in vision \cite{koh2020concept, losch2019interpretability, ismail2024concept, oikarinen2023label, laguna2024beyond, srivastava2024vlgcbm} and  more recently extended to natural language \cite{sun2024concept, nlp_concept_2}, where CBMs rely on concepts which are either human-annotated or extracted via multi-modal models s(e.g., CLIP) or LLMs. {Our work defined and used concepts in code.}

\paragraph{Learning Code Semantics for SE tasks.} 
Several approaches have been developed to learn code semantics for improving code reasoning tasks. Some models are inspired by domain algorithms~\cite{deepdfa}. Some work~\cite{semcoder,traced, fuzzpretrain,cycle, codeio, selfpico} conducted fine-tuning or pre-training on LLMs using different aspects of code semantics, e.g., execution data and high-level function description. For example, DeepDFA~\cite{deepdfa} uses dataflow embeddings with GNNs for vulnerability detection. 
SemCoder~\cite{semcoder} and TRACED~\cite{traced} leverage execution-derived signals; Cycle~\cite{cycle} iteratively improves generation using execution feedback. Our work learned semantics via supervised concepts.



\paragraph{Hierarchical Reasoning.} 
{Chain-of-thought prompting \cite{multi_step_f2} shows that eliciting explicit intermediate steps can improve arithmetic and logical reasoning. Subsequent prompting methods guide this process in different ways: least-to-most prompting \cite{multi_step_2} decomposes a problem and solves subproblems sequentially; decomposed prompting \cite{multi_step_1} solves each subtask with a separate prompt; Plan-and-Solve \cite{multi_step_3} requires an explicit plan before execution; and Self-Ask \cite{multi_step_4} generates and answers intermediate questions prior to the final answer. A separate line of work learns multi-step procedures from smaller self-generated or synthetic trajectories. STaR \cite{bootstrapping_3} and proxy-based bootstrapping \cite{boostrapping_2} iteratively generate, filter, and train on intermediate traces or proxy targets. Our work fine-tuned models first predicting code concepts and then code reasoning goals.

%% file: sections/conclusions.tex
\section{Conclusions and Future Work}
In this work, we proposed ConceptCoder, a novel concept learning framework aimed at improving large language models' performance on complex code reasoning tasks. By jointly learning fine-grained code concepts and reasoning objectives through multi-task learning, ConceptCoder enables models to better capture the semantic properties of code critical for reasoning. Applied to tasks such as vulnerability detection and branch prediction, ConceptCoder significantly improves both concept understanding and downstream task performance for all the models we experimented. It enhances model generalization and robustness and outperforms SOTA baselines.
These results demonstrate that concept-based fine-tuning is an effective method for addressing the critical challenge of code reasoning timely needed by current AI. In the future, we plan to extend concept learning to other software engineering tasks.



\section*{Impact Statement}
This work advances code reasoning capabilities in large language models through a novel fine-tuning method.  Our extensive evaluation demonstrated that this approach improved performance for all the LLMs we experimented and achieved the best performance among all the SOTA baselines we can find. The approach has a potential to extend to more code reasoning tasks

This research also brings the "concepts" concurrently shown useful in vision and NLP domains to software engineering, and demonstrated its usefulness beyond current intepretability applications.  Thus, our research potentially enables more future works on code concepts and concept-based fine-tuning in other domains. 



%% file: sections/appendix.tex
\section{Appendix}

\subsection{Data Collection and Processing for BP}
\label{app:app_bp_dataset}
We use TRACED’s tools~\cite{traced} to identify branch locations and determine ground truth labels (whether a branch is executed for a given input). We extended our static analysis tool to identify the concepts for BP. For example, to determine whether {\tt if (x > 0)} can be taken, we need to know current value of x. That value is typically set by earlier assignments, such as {\tt x = y + 1}.  Therefore, we will need to extract abstract values from this assignment.



To map each expression to an abstract value, we follow the prior work \cite{traced, trace-pradel}. As illustrated by the examples in Table \ref{tab:concept_map_bp}, conditional predicates are mapped to {\tt True} or {\tt False}. Arithmetic expressions are abstracted into range-based categories according to predefined numeric intervals, including {\tt zero}, {\tt Positive regular} (values between 1 and 1000), {\tt Positive large} (values greater than 1000), {\tt Negative regular} (values between $-1000$ and $-1$), and {\tt Negative large} (values less than $-1000$). For character assignments, if the assigned character is alphabetic (e.g., A--Z or a--z), it is labeled {\tt Alphabetic}; otherwise, it is classified as {\tt Non-alphabetic}. Pointer assignments are labeled either {\tt NULL} or {\tt Not NULL}. Finally, any assignment to an array index is considered {\tt Initialized}.

We use the CodeNet dataset \cite{codenet} which includes C language solutions to ~1,900 programming problems and test inputs. We follow the TRACED~\cite{traced} setup to generate traces. We sampled 52060 traces for training and evaluating ConceptCoder, while keeping the rest for the concept probing. We split the datasets into training, validation, and test sets with an 80:10:10 ratio of problems.

\subsection{Concept Probing Dataset}
\label{app:concept_probing_dataset}
    
To avoid data leakage when training our linear concept probing models, we use non-overlapping datasets for concept probing and for the reasoning task, but the datasets still contain the same set of concepts. 

For vulnerability detection (VD), we construct a concept-probing dataset of 28,974 examples, matching the size of the concept dataset.


For BP, we split the total trace dataset----52060(5/7) of the total used for training and evaluating ConceptCoder and the rest 20260(2/7) used for concept probing, as a linear model requires less data compared to fine-tune LLMs.

\subsection{Concept Probing for VD}
\label{concept_probing_performance}
\begin{figure*}[]
   \centering
    \resizebox{0.94\textwidth}{!}{ 
    \begin{subfigure}[t]{0.48\textwidth}
        \centering
        \includegraphics[width=\linewidth]{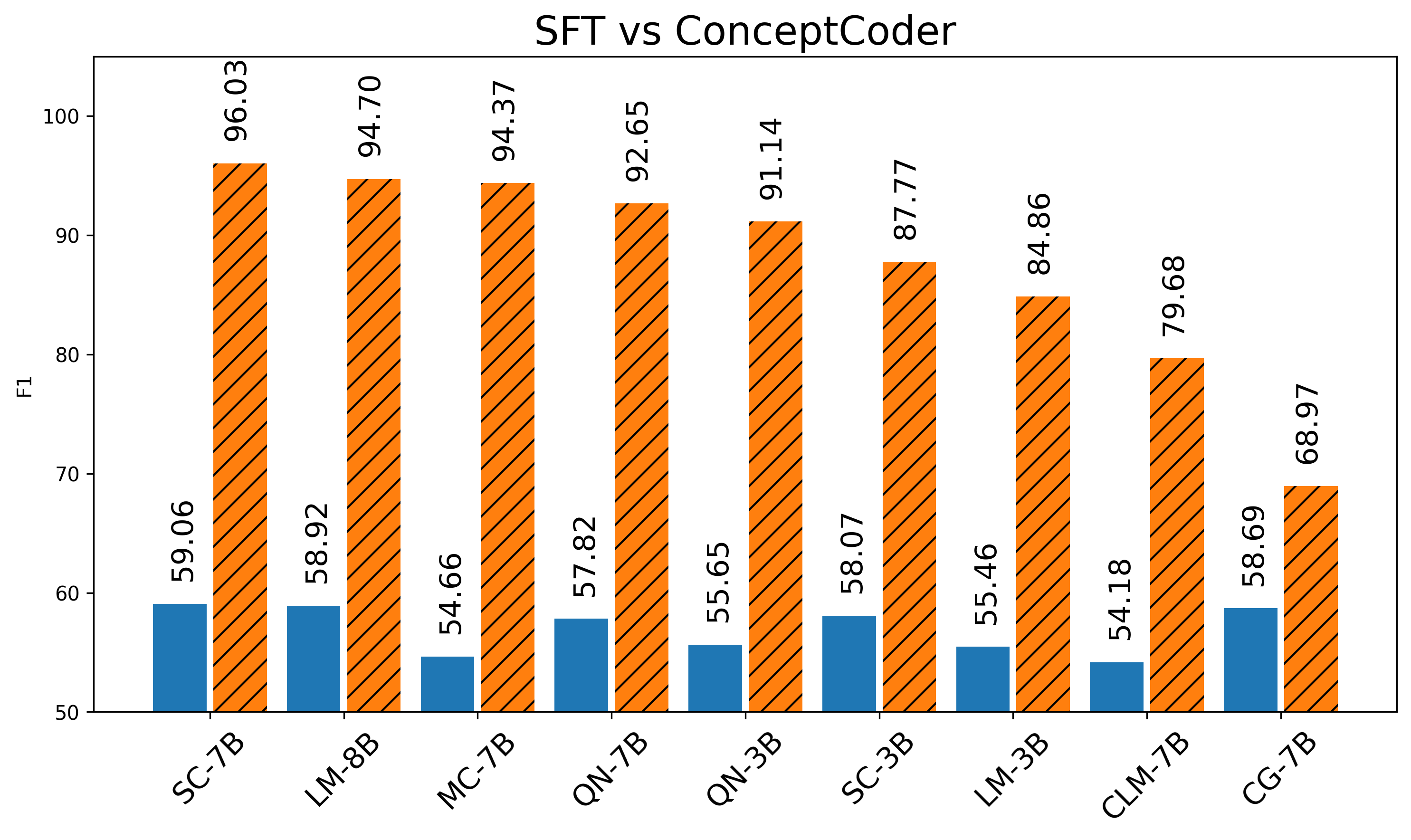}
        \caption{Concept probing performance on {\it concept} dataset.}
            \label{fig:rq1-result3}
    \end{subfigure}
    \hfill
    \begin{subfigure}[t]{0.48\textwidth}
        \centering
        \includegraphics[width=\linewidth]
        {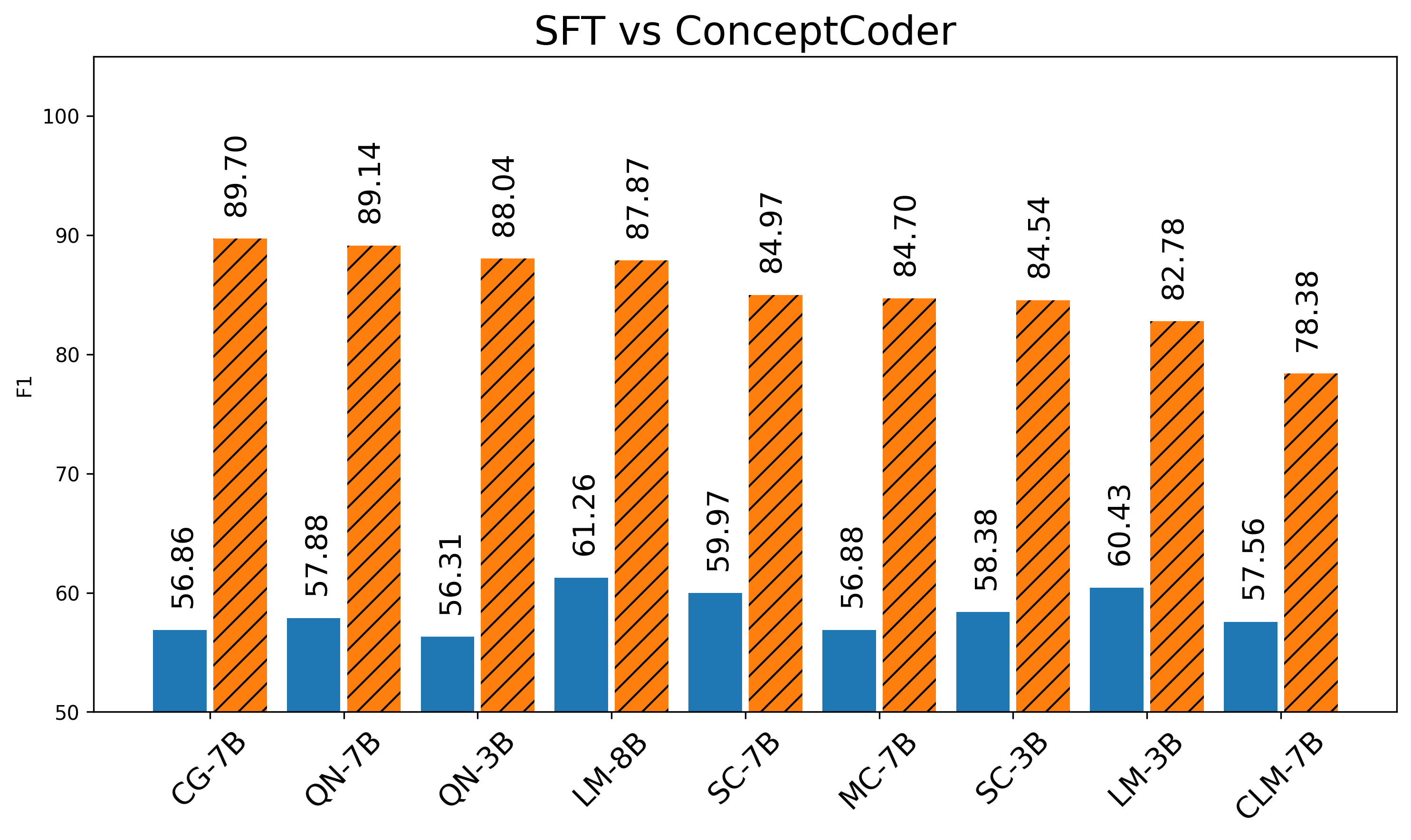}
        \caption{Concept probing performance on {\it general} dataset}
            \label{fig:rq1-result4}
    \end{subfigure}
    }

    \caption{ConceptCoder significantly outperformed SFT on concept recognition on both dataset.}
    \label{fig:vd_concept_recog}
    
\end{figure*}

Figure \ref{fig:vd_concept_recog} shows that ConceptCoder significantly outperform the concept recognition on both concept dataset and general dataset.

\newpage
\subsection{Hierarchical Prompting }
\label{cot_prompt}

Alongside our fine-tuning approach, we also investigated a prompt-based method to teach the model the target concepts at inference time. Listing~\ref{lst:prompt} presents our prompt design.

\begin{lstlisting}[style=prompt,caption={Hierarchical prompt with Concepts.},label={lst:prompt}]

You are a security researcher tasked with identifying vulnerabilities in a codebase. You have been given a function to analyze.  The function may or may not be vulnerable. So, at first, analyze the vulnerability related concepts first and then decide whether the function is vulnerable or not.

If you think it is vulnerable reply with @@VULNERABLE@@, otherwise reply with @@NOT VULNERABLE@@
    
For example:

@@VULNERABLE@@

Here is the function:
```
{target function}
```

Question: Does the function contain any statements related to the concept `Null Assignment`?
Answer: Yes, the function contains the following statements related to the concept `Null Assignment`: state->message = NULL

Question: Does the function contain any statements related to the concept `Null Check`?
Answer: No, the function does not contain any statements related to the concept `Null Check`

....

Question: Does the function contain any vulnerabilites?

Solve this problem step by step. Carefully break down the reasoning process to arrive at the correct solution. Explain your reasoning at each step before providing the final answer. The final answer should be either @@VULNERABLE@@ or @@NOT VULNERABLE@@.
    
\end{lstlisting}

\paragraph{Result.} 

Figure \ref{tab:hierarchical_prompt} presents the result for the two best proprietary models. The result indicates that hierarchical prompting does not help enough and fine-tuning is necessary for stronger performance.
\begin{table}[H]
    \centering
    \renewcommand{\arraystretch}{1.15}
    \caption{Hierarchical prompting performance by the two best proprietary models}
    \resizebox{0.48\textwidth}{!}{ 
    \begin{tabular}{l|c|c}
    \hline
        \textbf{Model} & \textbf{Concept Dataset} & \textbf{General dataset}  \\ \hline \hline
        GPT-5.2 & 32.35 & 28.12\\
        Claude-Opus-4.5 & 47.22 & 41.26\\
    \hline
    \end{tabular}
}
    \label{tab:hierarchical_prompt}
\end{table}